\documentclass[twocolumn,showpacs,preprintnumbers,amsmath,amssymb,nofootinbib]{revtex4}

\usepackage{graphicx}
\usepackage{dcolumn}
\usepackage{bm}

\begin{document}

\preprint{arXiv:0705.3884[hep-ph]}

\title{Inclusive electron spectrum in the region of pion production 
in electron-nucleus scattering and the effect of the quasi-elastic interaction}

\author{Hiroki Nakamura${}^1$, Tadaaki Nasu${}^2$, Makoto Sakuda$^2$}
\email[Correspondence: ]{sakuda@fphy.hep.okayama-u.ac.jp} \author{Omar Benhar${}^3$}

\affiliation{%
${}^1$Department of Physics, Waseda University, Tokyo 169-8555, Japan.\\
${}^2$Department of Physics, Okayama University, Okayama 700-8530, Japan\\
${}^3$INFN, Sezione di Roma and Dipartimento di Fisica, Universita "La Sapienza", 
1-00185 Roma, Italy\\
}%

\date{\today}

\begin{abstract}
We have carried out a calculation of the inclusive electron scattering cross section 
off oxygen in the kinematical region corresponding to beam energies between 700 and 1200 MeV, 
where quasielastic scattering and single pion production are the dominant reaction 
mechanisms. The formalism developed and successfully applied to describe quasielastic 
scattering has been extended to include both $\Delta$ production and non-resonant 
pion production. The results are in fairly good agreement with experimental data 
over the whole range of energy transfer, including the dip region between the quasielastic 
peak and the first resonance.
\end{abstract}

\pacs{21.60.-n,24.10.Cn,13.40.-f,25.30.Fj}
\maketitle

The precise determination of the neutrino cross section is vital to the
analysis of neutrino oscillation experiments \cite{Kajita}
and neutrino astrophysics \cite{SN}. 
For example, neutrino oscillations depend on the energy of
the parent neutrinos, which are estimated from the information on
the spectrum of the secondary particles. As a consequence, the study of neutrino oscillations
requires a theoretical model capable of providing an accurate description of the 
spectrum of charged leptons produced in nuclear weak interactions.
Recent experimental and theoretical developments in the field of 
neutrino-nucleus interactions are reviewed in the Proceedings of the NuInt Workshops
\cite{nuint}.

In a previous paper \cite{Benhar1} we have studied the inclusive electron cross 
sections off oxygen, at beam energies between 700 and 1200 MeV and scattering angle 
32$^\circ$, using a theoretical approach 
in which the initial state of the target nucleus is described by a realistic
spectral function and final state interactions (FSI) between the struck nucleon and
the spectator particles are consistently taken into account. The results of these
calculations agree with the data at $\sim$10\% level in the region of the 
quasielastic peak, while sizably underestimating the measured cross sections 
above pion production threshold. The authors of Ref. \cite{Benhar1} ascribed this 
problem to deficiencies in
the description of the nucleon structure functions in the $\Delta$ production 
region \cite{Benhar1,Benhar2}.

In the present paper, we apply the approach of Ref. \cite{Benhar1} and use the MAID
model \cite{MAID}, which includes both resonant and non-resonant contributions, 
to describe pion production. 
The results obtained from this model are in excellent agreement with 
the existing measurements of pion photo- and electro-production off nucleon target 
at energies up to 1 GeV.
The model contains the contributions of the Born term ($\pi NN$),
 $\rho$/$\omega$ exchange and the resonances, from the $\Delta$ to higher nucleon 
resonances. 
In order to clearly identify  nuclear effects we compare theoretical results 
to data for both hydrogen \cite{OConnel,Sealock,Niculescu} and oxygen \cite{data} 
targets.

First, we review the formalism of inclusive electron-nucleon scattering.
We consider the process $e + N \to e + X$, where $X$ is the undetected hadronic final state, 
which may contain multiple hadrons.
In the rest system of the target nucleon, the differential cross section can be written as
\begin{eqnarray}
\nonumber
d\sigma_{eN} & = & (2\pi)^4 \delta^4(p+k-p^\prime-k^\prime) \frac{1}{4E_e M} 
\overline{ \left| {\cal M}_{eN\to eX} \right|^2 } \\  
&  & \ \ \ \ \ \ \times \frac{ d^3 {\bm k}^\prime }{ (2\pi)^3 2E_e^\prime }
  \frac{ d^4 p^\prime }{ (2\pi)^3 2p_0^\prime }P_x( {\bm p}^\prime , p^\prime_0 ) \ ,
\label{eq:def}
\end{eqnarray}
where $k\equiv(E_e,0,0,E_e)$ $k^\prime\equiv(E^\prime_e,0,0,E^\prime_e)$, $p\equiv(M,0,0,0)$ and 
$p^\prime\equiv(p^\prime_0,{\bf p}^\prime)$ are the four-momenta of 
the incident electron, the scattered electron, the target nucleon and the hadronic final state, 
respectively.

The invariant amplitude ${\cal M}_{eN\to eX}$ includes electromagnetic interaction only, 
and does not contain hadronic final state interactions. 
If the final hadronic state $X$ consists of multiple hadrons, the invariant amplitude
is integrated over the relative four-momenta of the hadrons and depends only on the
their total four-momentum $p^\prime$.
After summing up and averaging over the spins and isospins of the initial and
final state particles, $\overline{\left| {\cal M}_{eN\to eX} \right|^2}$ is a function of 
the Mandelstam variables $s$, $t$ and $u$ and of the invariant mass 
$W=\sqrt{{p^\prime_0}^2-|{\bm p}^\prime|^2}$. 
The  function $P_x({\bm p}^\prime,p^\prime_0)$ describes the energy spectrum of the 
hadronic final state, 
In the case of elastic scattering, this state consists of a single nucleon
and $P_x({\bm p}^\prime,p^\prime_0) = \delta(p^\prime_0-\sqrt{{|\bm p}^\prime|^2+M^2})$,
while if the hadronic final  state is a $\Delta$ resonance, 
\begin{equation}
P_x({\bm p}^\prime,p^\prime_0)=f_{BW}(W)\frac{p^\prime_0}{W} \ ,
\label{eq:pxbw}
\end{equation}
where  $f_{BW}(W)$ is the Breit-Wigner function
\begin{equation}
f_{BW}(W) =\frac{1}{2\pi}\frac{\Gamma_\Delta}{(W-M_\Delta)^2+\Gamma_\Delta^2/4} \ .
\end{equation}
In the above equation $M_\Delta$ and $\Gamma_\Delta$ denote the $\Delta$ mass and width, 
respectively, while the factor $p^\prime_0/W$ accounts for the fact that the variable $p^\prime_0$ is
replaced by $W$, with $WdW=p^\prime_0 dp^\prime_0$.

From Eq.~(\ref{eq:def}) we obtain the doubly differential cross section 
\begin{eqnarray}
\nonumber
\frac{d\sigma_{eN}}{dE_e^\prime d\Omega} & = & 
\frac{1}{64\pi^2}\frac{E_e^\prime}{E_e} \frac{1}{M p^\prime_0} \\
 & & \ \ \times \overline{\left| {\cal M}_{eN\to eX} \right|^2} \ 
   \left.  P_x({\bm p}^\prime,p^\prime_0)\right|_{p\prime=p+q},
    \label{eq:difn}
\end{eqnarray}
where $q\equiv(\omega,{\bm q})=k-k^\prime$ is the 4-momentum transfer. 

Within the impulse approximation (IA) formalism \cite{Benhar1}, the generalization of the 
above result to the case of scattering off a {\em moving} nucleon carrying four momentum
$p \equiv (p_0,{\bf p})$, with $p_0=\sqrt{{|\bm p}|^2+M^2})$, 
can be used to obtain the electron-nucleus cross section through
\begin{eqnarray}
\nonumber
\frac{d\sigma_{eA}}{dE_e^\prime d\Omega} & = & \int d^3 p dE
    \frac{1}{64\pi^2}\frac{E^\prime_e}{E_e}\ \frac{1}{p_0 p^\prime_0}
    \overline{\left| {\cal M}_{eN\to eX} \right|^2} \\
    & & \left. \times
    P_h({\bm p}, E)
    P_x({\bm p}^\prime,p^\prime_0)\right|_{{p^\prime}={p}+{\widetilde{q}}},
    \label{eq:difa}
\end{eqnarray}
where the spectral function $P_h({\bm p},E)$ is the probability of
removing a nucleon of momentum ${\bm p}$ and energy $E$ \footnote{$E$ is the removal energy and 
is not the kinetic energy of the nucleon.}
from the target nucleus \cite{BFF}, normalized to the mass number $A$, and
$\widetilde{q} \equiv(\widetilde{\omega},{\bf q})$, with 
\begin{equation}
\widetilde{\omega} = p^\prime_0 - p_0 = \omega - E + M  - p_0  \ .
\label{def:omegatilde}
\end{equation}
Eqs.(\ref{eq:difa}) and (\ref{def:omegatilde}) show that, while in electron scattering off
a free nucleon the struck particle is given the entire four-momentum transfer, in a scattering 
process involving a bound nucleon a fraction $\delta \omega = \omega - \widetilde{\omega}$ 
of the energy transfer goes into the spectator system. The interpretation of 
$\delta \omega$ becomes particularly transparent in quasielastic scattering (corresponding to $W=M$) 
in the 
limit $|{\bf p}|/M \ll 1$, in which case Eq.(\ref{def:omegatilde}) yields $\delta \omega = E$. 

If the target is a nucleon at rest, the spectral function can be set to
 $P_h({\bm p}, E)=\delta^3({\bm p})\delta(E)$, implying $p_0 = M$ and 
$\widetilde{\omega} = \omega$, and Eq.(\ref{eq:difa}) reduces to Eq.(\ref{eq:difn}).

In our formalism, nuclear effects are described by the spectral functions $P_h({\bm p},E)$ and 
$P_x({\bm p^\prime},p^\prime_0)$. While, in general, $P_x({\bm p^\prime},p^\prime_0)$ can include the 
effects of FSI in the nucleus, if these effects are neglected it is the same as 
that of electron-nucleon scattering. 
For $P_h({\bm p},E)$ we use the results of the approach developed in Ref. \cite{LDA}, which includes 
the contribution of both quasiparticle states and the continuum arising from strong nucleon-nucleon 
(NN) correlations.

In the Fermi gas (FG) model the nucleon spectral functions appropriate to 
describe quasielastic scattering read
\begin{eqnarray}
\label{Ph:FG}
P_h({\bm p},E) & = & V \theta(p_F-|{\bm p}|)\delta(E+p_0-M-E_B) \ , \\
\label{Px:FG}
P_x({\bm p}^\prime,p^\prime_0) & = & \theta(|\bm p'|-p_F)
\delta(p^\prime_0-\sqrt{|{\bm p}^\prime|^2+M^2}) \ ,
\label{eq:fgpx}
\end{eqnarray}
where $p_F$ and $E_B$ denote the Fermi momentum and the average (negative) 
binding energy, respectively,
while $V$ is the normalization volume.
The step function in Eq.~(\ref{eq:fgpx}) takes care of Pauli blocking of the states belonging 
to the Fermi sea. In this paper, we have used $p_F=221$ MeV and $E_B=25$ MeV.

In the case of quasielastic scattering, and neglecting FSI, the cross section can be obtained from 
Eq.(\ref{eq:difa}) using $P_x({\bm p^\prime},p^\prime_0)$ of Eq.(\ref{Px:FG}) and setting
$p^\prime_0=\omega - E + M$. The effect of FSI can be included using the formalism discussed 
in \cite{Benhar1},  
yielding a quantitative account of the nuclear transparency measured in $(e,e^\prime)$ 
reactions \cite{Rohe}. 

Equation (\ref{eq:difa}) can be easily rewritten in such a way as to establish a 
relation between electron-nucleon ($eN$) and electron-nucleus ($eA$) cross sections. 
Neglecting again FSI, from Eqs.(\ref{eq:difn}) and (\ref{eq:difa}) we obtain
\begin{eqnarray}
\nonumber
\frac{d\sigma_{eA}}{dE^\prime_e d\Omega} & = & \int d^3p dE
 \frac{E^\prime_e}{E_e} \frac{1}{p_0} P_h({\bm p},E) \\
& & \ \ \ \ \ \ \ \ \ \ \ \times \left( \frac{M E_R}{E^\prime_R}  
\frac{d\sigma_{eN}}{dE^\prime_R d \Omega_R} \right) \ ,
\label{sigma_A}
\end{eqnarray}
where the subscript $R$ labels the electron kinematical variables in the 
frame in which the struck nucleon is at rest.
The expression in parentheses in the above equation is a function of
these variables, but is itself an invariant.
From the definition of the Mandelstam variables in the rest frame of the struck nucleon
we readily obtain
\begin{eqnarray}
E_R & = & \frac{s-M^2}{2M} \label{eq:er}\\
E^\prime_R & = & \frac{-u+M^2}{2M} \label{eq:erp}\\
\sin^2 \theta_R/2 & = & -\frac{t}{4 E_R E^\prime_R} \label{eq:tr}\ ,
\end{eqnarray}
where $s=(k+p)^2$, $t=(k-k^\prime)^2$ and $u=(k^\prime-p)^2$.

The MAID model \cite{MAID} provides a variety of cross sections and
amplitudes for pion production in electron-nucleon scattering. It
includes the $\Delta$(1232) and higher resonances, up to 2 GeV,
as well as the non-resonant contributions.
We have first employed the MAID model to calculate the pion production cross section
in $eN$ scattering. 

In the target rest frame the cross section can be written 
\begin{equation}
\frac{d\sigma_{eN}}{dE'_R d\Omega_R}
=\Gamma(\sigma_T+\epsilon\sigma_L),
\end{equation}
where
\begin{eqnarray}
\Gamma &=& \frac{\alpha}{2\pi^2}\frac{E'_R}{E_R}\frac{k_\gamma}{Q^2}
                                     \frac{1}{1-\epsilon} \\
\epsilon &=& \left(1+\frac{2{\bm q}^2}{Q^2}\tan^2\frac{\theta_R}{2}\right)^{-1} \\
k_\gamma &=& \frac{W^2-M^2}{2M} \ ,
\label{kgamma}
\end{eqnarray}
and $\sigma_T$ and $\sigma_L$ are provided by the MAID model as a function of $Q^2 $ 
and the invariant mass $W$, with $W^2=(p+q)^2$. 
We note that in Eqs.(\ref{sigma_A})-(\ref{kgamma}) we assumed the hadronic tensor  
$W^N_{\mu\nu}(p,q)$ to 
be $W^N_{\mu\nu}(p,q)=-W^N_1 g_{\mu\nu}+W^N_2 p_{\mu}p_{\nu}/M^2$, $W^N_1$ and $W^N_2$ being 
the nucleon structure function, instead of the one used in Ref.\cite{Benhar1}.
Using this somewhat simplified form appears to be reasonable, as for quasi-elastic scattering
the resulting cross sections are very close to those reported in Ref.\cite{Benhar1}.
\begin{figure}[tb]
\includegraphics[scale=0.4]{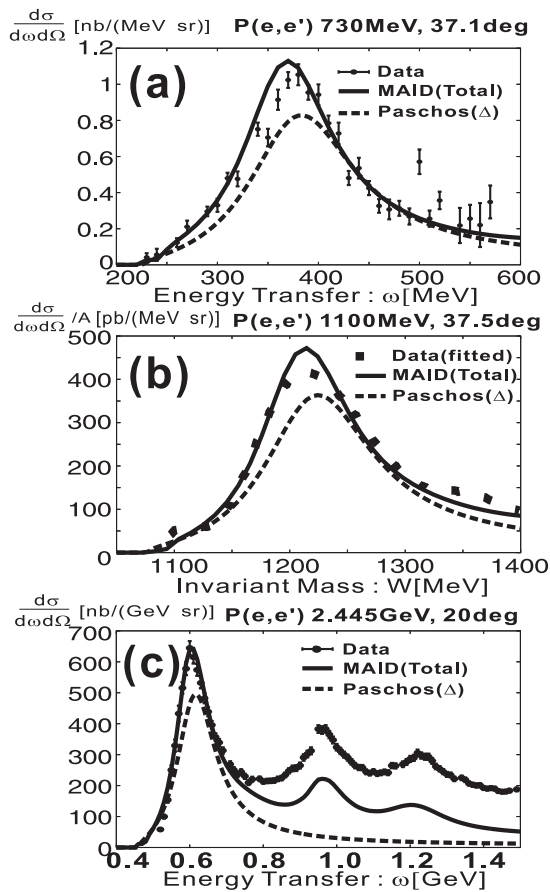}
\caption{\label{fig:oco}
Differential cross section of pion production
in electron-proton scattering, as a function of the energy transfer $\omega$
or the invariant mass $W$. The solid lines represent
the prediction of the MAID model, while the dashed lines
show the contribution of $\Delta$ production according to the model of
Paschos et al. \cite{Paschos}.
(a) beam energy $E_e=730$ MeV and the electron scattering angle
$\theta_e=37.1^\circ$ \cite{OConnel};
(b) $E_e=1100$ MeV and $\theta_e=37.5^\circ$ \cite{Sealock};
(c) $E_e=2445$ MeV and $\theta_e=20^\circ$ \cite{Niculescu}.  }
\end{figure}

In Fig. \ref{fig:oco}, we compare our results to the experimental data of 
Refs. \cite{OConnel, Sealock, Niculescu}.
The solid lines, obtained from the MAID model, include both 
resonance production and non-resonant pion production. 
In order to single out the contribution of $\Delta$ production we have also calculated
the cross section using the model of Paschos et al. \cite{Paschos}, with the updated
values of the parameters for the $P_{33}(1232)$ resonance given in Table I of
Ref. \cite{Olga}. The corresponding results are shown by the dashed lines.
The solid lines are in fairly good agreement with the data, although an excess of cross 
section, at the level of $\sim$5\%, is observed in some regions. The 
significant differences between the solid and dashed lines must be mainly ascribed 
to the contribution of non-resonant pion production, as Fig.4 of Ref.\cite{Olga}
has shown that the helicity amplitudes of the $\Delta$ resonance in the Paschos model and 
the MAID model are consistent with each other.
The interference between the $\Delta$ resonance and the non-resonant amplitude
can contribute to a part of this difference.
We also note that the peak of non-resonant pion production is shifted to lower 
energy transfer, with respect to that of the $\Delta$ resonance.  
Panel (c) shows that, while the agreement between calculations and data is good in 
the region of the $\Delta$ resonance, a significant deficit in the calculated 
cross section occurs at larger $\omega$, above $\sim 700$ MeV.
As the MAID model includes the contributions of the second and third resonances, 
the excess of the measured cross section is likely to be due to 
the deep-inelastic scattering and the onset of the multi-pion production,
 not taken into account in the model.
\begin{figure}[tb]
\includegraphics[scale=0.3]{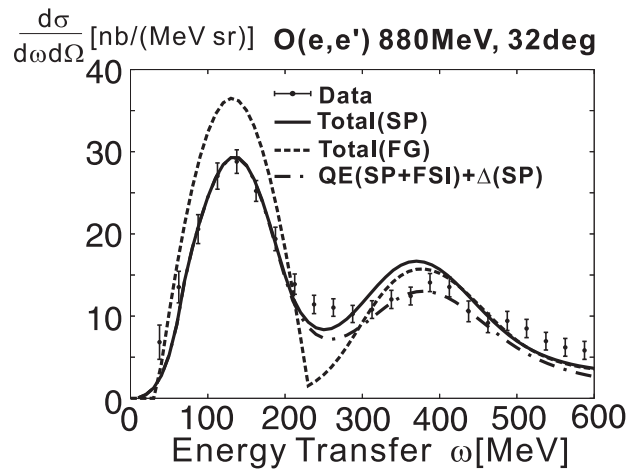}
\caption{\label{fig:o880} Differential cross section for $^{16}O(e,e')$ scattering
at beam energy $E_e=880$ MeV and scattering angle $\theta_e=32^\circ$.
The solid line shows the results obtained combining the quasielastic cross section
obtained from the approach of Ref. \cite{Benhar1} and the pion production 
cross section computed using the same spectral function and the MAID model. 
The dash-dot line represents the results obtained including $\Delta$ production
only, according to the model of Ref. \cite{Paschos}. For comparison, the dashed line
shows the cross section predicted by the FG model. The experimental data are taken 
from Ref. \cite{data}.}
\end{figure}

In Fig.~\ref{fig:o880} we show the comparison between calculations and data 
for the $^{16}O(e,e')$ cross section at beam energy 880 MeV and scattering angle 32$^\circ$, 
as a function of the energy transfer $\omega$.
The calculated cross section includes both quasielastic and pion production channels,
the latter being described according to the MAID model, under the
assumption that the electron-proton and electron-neutron cross sections be the same. 
The solid line has been obtained using the spectral function
of Ref. \cite{LDA} to model the momentum and energy distribution of the struck nucleon. 
The effect of FSI in quasielastic scattering has been also included, following the 
approach described in Ref. \cite{Benhar1}.
The results agree reasonably well with experimental data over the whole range 
of energy transfer. 
For comparison we also plot, by the dash-dot line, the sum of the 
quasielastic cross section  and the contribution of $\Delta$ production only, obtained from 
the same spectral function using the model of Ref. \cite{Paschos}.
The non-resonant pion production cross section, which corresponds to the difference 
between the solid and the dash-dot lines, turns out to be appreciable in the dip region 
between the quasielastic peak and the resonance bump.
We note that quasielastic scattering also contributes a significant amount of strength
to the dip region, due to the presence of sizable high-momentum and high-energy components
in the spectral function and to the effect of FSI. 
To illustrate the importance of using a realistic spectral function we show, by the dashed
line, the results obtained from the FG model (Eqs.(\ref{Ph:FG}) and (\ref{Px:FG})).
It clearly appears that this model overestimates the measured cross section 
in the region of the quasielastic by $\sim20$\%, while a 
large deficit is observed in the dip region. 
\begin{figure}[t]
\includegraphics[scale=0.3]{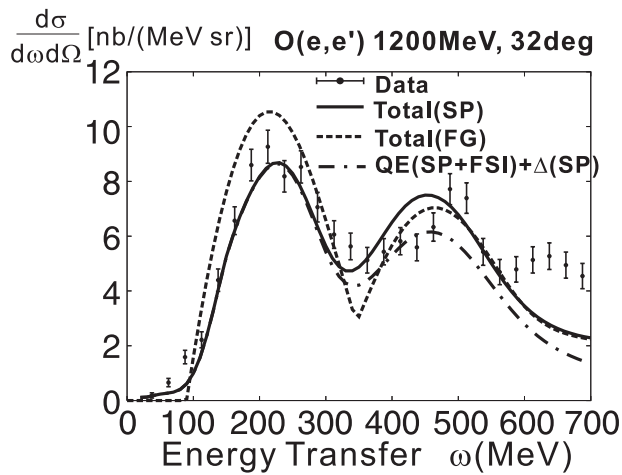}
\caption{\label{fig:multi}
Same as in Fig. \ref{fig:o880}, but for beam energy $E_e=1200$ MeV.  }
\end{figure}

In Fig.~\ref{fig:multi}, we plot the same cross sections as in Fig. \ref{fig:o880}, 
but for beam energy $E_e=1200$ MeV.
The main features of the results are the same. In both cases, the agreement between 
the calculations based on our model and the data is fairly good. We note that 
we observe a similar deficit in
the calculation in the region of higher energy transfer, $\omega>600$ MeV, to Fig.1(c). 

In conclusion, we have supplemented the quasielastic scattering calculation of 
Ref. \cite{Benhar1}, based on the use of a realistic nuclear spectral functions 
and including the effects of FSI, with the MAID model of pion production. 
The resulting model provides a reasonable description not only of the quasielastic 
and resonance regions, but also of the region of the dip. 
The analysis of the different contributions to the inclusive cross section 
suggests that the strength in the dip region receives contributions from both 
non-resonant pion production and quasielastic processes, due to the combined 
effects of the high-momentum and high-energy components of the spectral function 
and FSI. \\

Finally, we comment that our description of electron-nucleus scattering does not involve any 
adjustable parameters. The elementary cross sections employed in the 
calculations are extracted from electron-nucleon data, while the spectral 
functions are obtained from a nuclear hamiltonian which is fully
determined from the properties of the exactly solvable two- and
three-nucleon systems. Alternate many-body approaches, based on somewhat
oversimplified interaction models \cite{Gil}, also provide a fairly good 
description of the electron-nucleus cross sections at $Q^2 \sim 3$\ GeV$^2$.
However, while being more flexible and allowing for a consistent 
inclusion of some mechanisms not taken into account in our work, like,
e.g., long range correlations and the broadening of the $\Delta$ peak in
the nuclear medium \cite{Mosel,Singh}, these approaches rely on a
dynamical input that is not constrained by nucleon-nucleon scattering
data.\\

Within our formalism, the effect of FSI in the inelastic channels, 
leading to the broadening of the $\Delta$ peak, can be included through 
the folding procedure described in Ref.\cite{Benhar1}. However, the 
calculation of the folding functions associated with inelastic channels 
involves additional difficulties, and has not been carried out yet.
To estimate the relevance FSI effects, the authors of Ref.\cite{Benhar2}
have folded the inelastic cross sections using the same folding functions 
employed for the quasielastic channel. The results show that the main
effect is a quenching of the peak of less than $\sim 4$\% in carbon at 
$Q^2 \sim$ 0.4 GeV$^2$.\\

We thank T.Sato, C.Keppel and E.Christy for 
much advice. This work was supported in part by 
the Grants-In-Aid for the Japan Society for Promotion of Science (No. 18340066)
and the 21COE program at Waseda University. 
We note that a similar calculation using MAID model 
was also done by O.Buss et al.\cite {Mosel}.


\begin{thebibliography}{99}%
\bibitem{Kajita}
T. Kajita, Nucl.~Phys.~B (Proc.~Suppl.) {\bf 159}, 15 (2006);
M.H. Ahn et al. (K2K Collab.), Phys.~Rev.~D {\bf 74}, 072003 (2006);
D.G. Michael et al. (MINOS Collab.), Phys.~Rev.~Lett. {\bf 97}, 191801 (2006).
\bibitem{SN}
K.~Sumiyoshi, Nucl.~Phys.~B (Proc.~Suppl.) {\bf 159}, 27 (2006).
\bibitem{nuint}
{\it Proceedings of The Third International Workshop on Neutrino-Nucleus
Interactions in the Few-GeV
Region (NuInt04)}, Edited by F.Cavanna, P.Lipari, C.Keppel and
M.Sakuda, Nucl. Phys. B (Proc. Suppl.) {\bf 139} (2005);
{\it Proceedings of The Fourth International Workshop on Neutrino-Nucleus
Interactions in the Few-GeV Region (NuInt05)}, Edited by F.Cavanna, J.Morfin and
T.Nakaya, Nucl. Phys. B (Proc. Suppl.) {\bf 159} (2006).
\bibitem{Benhar1}
O.Benhar, N.Farina, H.Nakamura, M.Sakuda and R. Seki, Phys.~Rev.{\bf D72},
053005 (2005).
\bibitem{Benhar2}
O. Benhar and D. Meloni, Phys.~Rev.~Lett.~{\bf 97}, 192301(2006).
\bibitem{MAID}
D. Drechsel, O. Hanstein, S.S. Kamalov and L. Tiator, Nucl.~Phys.{\bf A645}, 145
(1999); The cross sections resulting from the MAID2003 model can be calculated 
at: http://www.kph.uni-mainz.de/MAID/maid2003/.
\bibitem{OConnel}
J.S. O'Connel et al., Phys.~Rev.{\bf C35}, 1063 (1987).
\bibitem{Sealock}
R.M. Sealock et al., Phys.~Rev.~Lett.~{\bf 62}, 1350(1989).
\bibitem{Niculescu}
I.~Niculescu et al., Phys.~Rev.~Lett.~{\bf 85}, 1186(2000).
\bibitem{data}
M.~Anghinolfi {\it et al.}, Nucl.\ Phys.\ {\bf A602}, 405(1996).
\bibitem{BFF}
O.~Benhar, A.~Fabrocini and S.~Fantoni, Nucl. Phys.{\bf A505}, 267(1989).
\bibitem{LDA}
O.~Benhar, A.~Fabrocini, S.~Fantoni and I. Sick, Nucl. Phys.{\bf A579}, 493(1994).
\bibitem{Rohe}
D.~Rohe et al., Phys.~Rev.~C {\bf 72}, 054602(2005).
\bibitem{Paschos}
E.A.Paschos, Ji-Y.~Yu and M.~Sakuda, Phys.~Rev.{\bf D69}, 014013(2004).
\bibitem{Olga}
O.Lalakulich, E.A.Paschos, and G.Piranishvili, Phys.~Rev.{\bf D74}, 014009(2006).
\bibitem{Gil}
A.Gil, J.Nieves and E.Oset, Nucl.Phys.{\bf A627}, 543(1997).
\bibitem{Mosel}
O.Buss, T.Leitner, U.Mosel and L.Alvarez-Ruso, Phys.~Rev.{\bf C76}, 035502(2007).
\bibitem{Singh}
S.K. Singh, M.J.Vincente-Vacas and E.Oset, Phys.Lett.{\bf B416}, 23(1998).
\end{thebibliography}
\end{document}